Szilvia Borbély[1] – Csaba Makó[2] – Miklós Illéssy[3] -Saeed Nostrabadi[4]

# Trade Union Strategies towards Platform Workers: Exploration Instead of Action*

(The Case of Hungarian Trade Unions)

[1] Trade Union Expert, PhD in Economics, email: h10095bor@ella.hu

[2] Prof.em.Dr., Public Service University, Budapest, email: Mako.Csaba@tk.mta.hu

[3] Senior Research, PhD, E.L. Research Network, Centre for Social Sciences, email: Illessy.Miklos@tk.mta.hu

[4] PhD Student. Szent István University, Phd School of Regional and Business Admninistration

**Contents**                                                              page

**List of Abbreviation**

**LIGA** = Democratic League of Independent Union - in Hungarian: Szakszervezetek Demokratikus Ligája (National Confederation)

**LIGA VISZ** = LIGA Hotel Industry and Catering – in Hungarian: LIGA Szállodaipari és Vendéglátóipari Szakszervezet (Sector level National Federation)

**MASZSZ** = Hungarian Trade Union Association – in Hungarian: Magyar Szakszervezeti Szövetség

**MASZSZ KASZ** = Hungarian Trade Union Association – Retail Sector Employees Union - in Hungarian: Magyar Szakszervezeti Szövetség – Kereskedelmi Alkalmazottak Szakszervezete (Sector level National Federation)

**MOSZ** = National Alliance of Workers Council – in Hungarian: Munkástanácsok Országos Szövetsége, (National Confederation)

NGTT = National Economic and Social Council – in Hungarian: Nemzeti Gazdasági és Társadalmi Tanács

**SZEF** = Co-operaton Forums of the Trade Unions – in Hungarian: Szakszervezetek Együttműködési Fóruma (National Confederation)

**ÉSZT** = Trade Union Association of Intellectuals – in Hungarian: Értelmiségi Szakszervezeti Tömörülés (National Confederation)

**C.A**. = Collective Agreement – in Hungarian: Kollektív Szerződés (KSZ)

**NIRC** = National Council of Interest Reconciliation – in Hungarian: Országos Érdekegyeztető Tanács - OÉT (national tri-partite body until 2011)

**NGTT** = National Economic and Social Council – in Hungarian: Nemzeti Gazdasági és Társadalmi Tanács, from 2011)

**VKF** = Standing Consultative Forum for the Private (literally: 'competitive') Sector = in Hungarian: Versenyszféra és a Kormány Állandó Konzultációs Fóruma - VKF)

**MKDSZ** = Labour Mediation and Arbitration Service – in Hungarian: Munkaügyi Közvetítő és Döntőbirói Szolgálat (Established in the middle of 1990's and re-established in 2017)
**SSDC** = Sectoral Social Dialogue Committees (Ágazati Párbeszéd Bizottság, ÁPB) (Supporting by the EU PHARE Program, 30 have been established since 2004)
**KATA** = Itemized Tax for Small Businesses – in Hungarian: Kisadózó Vállalkozások Tételes Adója

**SME** = Small and Medium Sized Enterprises – Hungarian: Kis- és Közepes Vállalatok, KKV) Hivatal, KSH)

# 1. Introduction: Fragmented and divided trade unions and the emergence of the Digital Worker

In recent years, the emergence of a wide range of digital platforms ranging from AirBnB in hospitality to Taxify in transportation, to a wide range of other online platforms of businesses large and small have transformed the labour market throughout the world. Here we focus specifically on Hungary and the response (or lack thereof) of the once powerful trade unions, who have struggled to adapt to this new dimension to the labour market.

Following the failure of the state-socialist political – economic regime, the Hungarian trade unions structure shifted from a monolithic to a more pluralistic one, moving from relative strength to weakness as advocates for workers' rights. Fragmented and divided trade union structure emerged with competing unions both at workplace and national levels during the first decade of the 1990's. (Makó, 1997). The outcome of these developments: today's trade union confederations/federations have become weak social partners in labour relations system of the country. The five national level, cross-sectoral trade union confederations are the following: Democratic League of Independent Trade Unions (LIGA), Hungarian Trade Union Association (MASZSZ), National Alliance of Workers Councils (MOSZ), Co-operative Forum of Trade Unions (SZEF) and Trade Union Association of Intellectuals (ÉSZT). Almost all confederations/ federations emphasize in their statutes political autonomy and (party) neutrality.

- The only exception is the Workers' Council (MOSZ), which explicitly represents Christian-democratic/Christian-socialist values.
- MASZSZ was created as the result of a merger in 2015 between two confederations, National Confederation of Hungarian Trade Unions (MSZOSZ) and Autonomous Trade Unions (ASZSZ).
- LIGA, MASZSZ and MOSZ are present mostly in the competitive sectors (industry and services). LIGA includes 95 trade unions with 100 000 unionised workers. MASZSZ has 31 member federations and around 104 000 union members. MOSZ covers 53 000 union members and composed by around 100 unions.
- ÉSZT covers research institutes and higher education, "intellectual workers", meanwhile SZEF covers first of all, primary, secondary and vocational education sector, local and central government, social services and partly health sector employees. SZEF has 14 affiliated unions, including one of the teachers' unions (PSZ); ÉSZT has eight affiliated unions, including employees in higher education and research.

FIDESZ in alliance with the Christian Democratic People's Party was elected in 2010 with a two-thirds majority in the Parliament, meaning a political turn in concern of industrial relations. Beginning in 2012, a strong decline of the social dialogue began to accelerate, and as a result of the change of the previous bodies, the unions' division widened and the strength of social dialogue was significantly weakened. In the private sector, the social dialogue became formal, with existing trade unions and employers' organisations remaining without meaningful power when it came to taking part in the decision-making process of workers' rights.

In 2011 the former – institutionalised –tripartite social dialogue at national level finished to function. In June 2011 the Parliament voted to replace the tripartite National Council for the Reconciliation of Interests (OÉT) by the National Economic and Social Council (NGTT), with the new body no longer included representation from the state. Its members include employers' and employees' organizations, NGOs, chambers and churches, but lack decision-making rights, retaining only the right to draft proposals to government.

In December 2011, under the pressure of both trade unions and employers' organizations, the government established a new permanent consultation forum-- the Permanent Consultation Forum (VKF)--with the participation of government and trade union confederations and employers' organisations, but only for the private sector. It was set up to discuss employment issues on the initiative of the social partners. However, only three trade union confederations and three employers' organisations have been invited to participate in this new body, and its role has been more limited than the role and rights of former OÉT. For example, until 2011, the government fixed the mandatory minimum wage and the so-called guaranteed wage minimum in a decree based on the results of the consultation with the National Council for the Reconciliation of Interests.

## 1.1 Continuous membership decline. Low collective agreement coverage rate and dominance of the company level bargaining (micro-corporatism)

Trade unions lost the majority of their members within a few years after the economic restructuring and mass privatisation during early 1990s. While under the state-socialist political and economic system union membership was almost compulsory, overall trade union density has now fallen to around 8 per cent (in 1998 it was 27,7%, in 2000 it was 9,4%, in 2015 it was 7,9% and in 2018 it was 7,9% (Source: OECD). According to the Hungarian National Statistical Office (HNSO), the highest level union density characterise the electric energy sector, followed by majority state owned sectors (i.e. public service like transport, education, health and social care). However, even among the state owned sector the decline among trade union membership decline significant, with the share of union membership being halved and varies between 15% and 22%. (Neumann, 2018).

In 2000, the collective bargaining (C.A.) coverage (the percentage of employees covered by collective agreement) was 47%; it has dropped by 2020 to 30%. The bargaining system currently is decentralized, uncoordinated with limited impact on working conditions, confined principally to single-employer agreements in the private sector and public companies. The majority of collective agreements (C.A.) are signed in large or medium-sized companies, with the coverage rate being highest among state- and municipality-owned companies. Collective agreements are low among SMEs, while the coverage rate is the highest in the group of the state- and municipality owned enterprises;36.7% of workforce covered by single-employer's agreement. As concerns the competitive sector, 21.8% of the work-force are covered by single employer's C.A., while 10.5% by multi-employer C.A.[5]

The number and coverage rate of different types of collective agreements by sector is presented below on the basis of the date collected by Borbély and Neumann (2019). As the authors noted: "These figures clearly indicate that the single-employer level is the dominant one. The majority of C.A. is signed in large or medium-sized companies. Coverage is highest among state- and municipality owned companies. There are even fewer genuine sectoral/industry agreements. Although the registry includes 19 valid industrial agreements in the private sector and one in the public sector, if the data are further scrutinised only five, covering different segments of the electricity industry and road transport, have been concluded or modified since 2011. New initiatives for industrial agreements are quite rare, the most notable exception being the health-care agreement signed in 2017." (Borbély–Neumann, 2019:302)

**1: Number and Coverage Rate of C.A. in Hungary (August 2017)**

| Types of C.A. | Sector | No. of C.A. | No. of companies, institutions | No. of employees covered by C.A. | Total no. of employees | Rate of C.A. coverage (%) |
|---|---|---|---|---|---|---|
| Single employers | Competitive sector* | 972 | 972 | 443 691 | 2 031 700 | 21.8 |
| | Budgetary sector | 1 629 | 1 629 | 259 887 | 707 500 | 36.7 |
| Multi - employer | Competitive sector | 66 | 3 621 | 214 262 | 2 031 700 | 10.5 |
| | Budgetary sector** | 1 | 3 | 320 | 707 500 | 0.1 |

---

[5] http://www.mkir.gov.hu/index.php. For total number of employees (at employers with at least five employees, without public works, July/2017): HCSO, https://www.ksh.hu/docs/hun/xstadat/xstadat_evkozi/e_qli033.html, https://www.ksh.hu/docs/hun/xstadat/xstadat_evkozi/e_qli034.html,

| | | | | | |
|---|---|---|---|---|---|
| **Total** | | 2 668 | - | 812 386 | 2 739 200 | 29.6 |

*Notes:*
\* The "competitive sector' includes private sector and state/municipality owned enterprises.
\*\*The nee sectoral health-care agreement is not included.
\*\*\*Single and multi-employer agreements' coverage should not be added up due to the overlap between the bargaining levels.
*Sources:* For collective bargaining: Collective agreement registry, Ministry of National Economy. For total number of employees (at employers with at least five employees, without public works, July/2017: Hungarian National Statistical Office (HNSO), In: Borbély – Neumann, 2019:303)

Sectoral agreements in Hungary used to be quite rare. Sectoral-level negotiation inhibited by the absence of employers' organizations prepared to negotiate. Although industry-level business associations exist, in most cases their role is exhausted in lobbying. Fierce competition among private companies within the same sector also hinders the development of sectoral bargaining. On the trade union side, industry federations are badly funded and staffed, and are unable to mobilize employees in the entire sectors. Trade unionists, however, would very much welcome sectoral collective bargaining, from which employees benefit, especially in SMEs and family-owned business sector where union organization and local bargaining are missing completely.

Since 1992 the Labour Code has allowed the use of an extension mechanism, but it has been used only in a few sectors: construction, hospitality (hotels and catering), electricity and baking. The 2012 new Labour Code curbed the rights and operating conditions of trade unions at the workplace and increased the scope of unilateral employers' decisions. Most collective agreements do not include terms covering the remuneration of employees, which increasingly is subject to influence from minimum wage legislation. Trade unions' primary responsibility is to develop a broad regulation framework of working conditions. In the private sector it is very rare that a C.A. includes wage scales, the so-called 'tariffs', which are supposed to be applied in determining individual basic wages. It is a general problem that a large proportion of C.A.s simply copy and paste regulations from the Labour Code. Only a small proportion of C.A.s contains meaningful stipulations on relations between the signatories, such as working time schedules, wage supplements and employment conditions. In Hungary, the procedural terms of C.A.s customarily include detailed regulation of the bargaining process, such as timing, negotiation rules, ratification's procedures, date of entry into force, termination and renegotiation. Other elements include coverage and time horizon and, in general, cooperation between the contracting parties. More generally, C.A.s may include topics related to industrial relations within the company or the industry, such as the rights and duties of employee representatives, the method of confirming the number of trade union members, the rights of trade union representatives, including their legal protection, arrangements for time-off for union work, access to an office and other infrastructure issues.

All C.A.s include substantive regulations setting the terms of employment, such as rules for hiring and firing, including the probation period, cases of immediate termination, duration of notice period and severance pay. Another important chapter of all C.A.s regulates conditions of employment, such as work schedules, working time, breaks during work, rest time, overtime, reference period for working time banks and the allocation of annual paid leave. In many sectors it is important that the C.A. contains rules of liability of fines for issues including inventory losses and damages arising from staff negligence, as well as employer's liability in cases of breaches of duty. The level of wage increases, and sometimes the basic wage in a wage scale, is regulated by a separate wage agreement that in most cases is signed annually. The annual agreement includes the in-kind part of compensation, too, including 'cafeteria' benefits, a form of flexible benefit system within which the employee can choose from a menu of possible benefits. Research has shown that Hungarian bargaining parties are reluctant to broaden the scope of bargaining (Bobély&Neumann, 2019). The reasons for the cautious behaviour of management in a couple of companies are manifold, from trying to avoid further conflicts following to the tension of past restructurings up to a corporate culture including the tradition of cooperative industrial relations. (Nacsa and Neumann, 2013 106)

## 2. Methodology and data collection

In order to understand and interpret the trade union leaders' attitudes towards digital labour in general and especially to platform work practice and employment, the CrowdWrok21 international research consortium decided to use the qualitative research method. This is: "An umbrella term covering an array of interpretive techniques which seek to describe, decode, translate, and otherwise come to terms with the meaning not the frequency, of certain more or less naturally occurring phenomena in the social world." (Maanen, 1979, In: Tomory, 2014:59). From the variety of qualitative research technics, the research tool chosen is the case study method. This research technique helps us to understand how people are thinking and interpreting their experiences in relation to the interest representation. An additional factor shaping this decision was that labour (industrial) relations of the digital labour and labour market are relatively under-researched in Europe, especially in the Central and Eastern Europe.

Our case study methodology comprises the use of two primary research techniques. First, we extensively deployed the tool of secondary analysis of the available documents reflecting the unions' position to the various forms of precarious employment with special attention to the platform work in the Central European region (Meszman, 2018, Sedlakova, 2018). Second, each national team members of the CrowdWork21 project countries (Hungary, Germany, Portugal and Spain) used a commonly designed and tested template of trade union questionnaire. (See in Annex no. 1)

The Hungarian team – during June – July 2020 – conducted eight interviews (using the computer aided telephoning, CATI – due to the COVID-19 pandemic) with trade union confederations/federations' leaders, officials and labour lawyers working closely with trade unions on platform work.

**Table 2: Trade union leaders, officials and experts interviewed on platform work (June – July 2020)**

| Job title | Date | Abbreviation of the trade union organisation | Trade union | Sector of activity |
|---|---|---|---|---|
| Head of Trade Union Secretariat | 2020.06.29, 10 -11 | MOSZ (1) | National Confederation | Mainly private sector |
| Trade union legal expert (labour lawyer) | 2020.06.29, 10 -11 | MOSZ (2) | National Confederation | Mainly private sector |
| PhD student (labour lawyer) | 2020.06.29, 10 -11 | MOSZ (3) | | |
| Union official: Responsible for union communication and training and industrial relations' expert | 2020.06.23. 10-11 | Liga | National Confederation | Mainly private sector |
| Union President | 2020.06.23.17.36 | Liga VISZ | Sectoral Federation, (member of LIGA) | Hotel industry and catering |
| Union organizational secretary | 2020.07.09. 10-11 | SZEF | National confederation | Mainly public sector |
| Union national coordinator and county secretary of Budapest and Pest county | 2020.06.29.11.06 | MASZSZ KASZ | Sectoral federation, (member of MSZSZ) | Retail sector |
| Union President | 2020.06.29.15.30 | ÉSZT | National Confederation | Mainly public sector |

# 3. Trade unions and platform work: Dominance of the 'wait and see' attitude

**3.1 Contradicting views on platform economy. Lack of systematic first hand experiences**

The views on digital labour/platform work diversified greatly by trade union federations ranging from "it is synonym of the restructuration of economy" (LIGA); "it is a novel economic form (ÉSZT); to "it is unavoidable, however, we cannot predict the future effect of the COVID-19 pandemic on it"(LIGA VISZ), to "it exists, develops and we have to live with it" (SZEF). We can expect its mass appearance in the retail sector, if the large transnational companies (e.g. TESCO, AUCHAN, ALDI, LIDL etc.) see potential in it (in case of further spread of on-line orders) and build up easily available solutions for the clients (MASZSZ KASZ). In the opinion of MOSZ (2), terms like "platform economy", "crowd work", "sharing economy", etc. have become fashionable, yet their use is conceptually chaotic. These terms are not generally used at all by SZEF, or MASZSZ - KASZ. ÉSZT concentrates on the content of the term. In the case of LIGA – VISZ, operating in the hospitality industry where, before the COVID crisis AirBnB did create a strong competition with the traditional hotels, this union recognized that the main source of the extremely high growth of AirBnB platform could be explained by their ability to exploit holes in the regulation and through unfair competition. "Whereas professional hotels must comply with security norms or professional standards, a household renting its apartment via a platform does not incur such compliance costs. The supposed higher 'efficiency' of such platform firms is thus a consequence of their ability to avoid regulations and they win market share from 'conservative' over-protected firms …" (Montalban – Frigant – Jullien, 2019:5) LIGA considers platform workers as an outsourced employee, for whom the trade union can provide legal aid and advocacy.

The interviewed trade unions usually follow the development of the platform economy, but this attitude was mostly shaped by their direct interest. The general concerns of trade unions on platform work fall into several categories:

1: They are neutral, necessary and unavoidable (LIGA, SZEF, ÉSZT).

2: They are scary, although in many cases they offer good solution (LIGA VISZ).

3: They are economically concerning for the large traditional companies but socially useful (MASZSZ-KASZ).

4: Their positive and negative features appear together (MOSZ). Positive features are that new forms of organization and logic appear, many people can quickly and flexibly get work, employment rates improve, young people do not increase the size of unemployed due to be involved in platform work. Negative features: platform firms seduce workers from the traditional companies and deteriorate stability of social security system.

5: We need to address them from the trade union point of view (ÉSZT).

6: We need to learn to live together and regulate them for the benefit of workers. A standard regulations system must be developed that protects both the interests of service providers (platform workers) and clients (SZEF).

## 3.2 Platform workers and the trade unions: Organising them is a low priority

It is revealing that the trade unions surveyed for this study currently *have no platform workers among their members.* Why unions do not prioritize this emerging and growing digital workforce can be explained in part by the fact that platform workers do not belong to the group of traditional employees they serve and it is sometimes difficult to to identify their employers. The workers typically define themselves (or are defined by the platform) as individual entrepreneurs, self-employed, freelancers or contractual workers. Frequently, they are participating in the 'grey zone of regulation' and their relations with the platform are not always transparent. There is also the sentiment that platform workers do not organize (yet) and are not interested in being union member. Even if they approach unions asking for some information, they typically do not wish to join (LIGA, LIGA – VISZ). Some union confederations insist that they are simple invisible to the trade unions and out of reach of the traditional recruitment and organizing technics. (LIGA, MOSZ). In Hungary, labor law regulation of platform work is missing and their position in the trade union is incomprehensible or nonexistent (MOSZ 1). In this relation we have to note that trade unions generally do not pay too much attention to atypical workers (MOSZ 3) as there is a long-standing reserved attitude towards atypical workers and precariat among the trade unions leaders in Hungary (Neumann, 2018).

According to the public sector trade unions, the job structure in this sector does fit for the characteristics of platform work (SZEF, ÉSZT).

All interviewed trade union officials agreed that 'class-consciousness' is missing among platform workers, but according to LIGA's representative, in Hungary even unionized workers have no 'class-consciousness' in comparison with Spain and Portugal. In addition, as MOSZ (1) head of secretariat noticed, there is no social-cohesion among the platform workers which could generate a kind of political activity in the future, and it is not certain that platform workers would even need any union-based or similar interest representative structure to get some help in their activities or to meet after work to socialize. However, there are already signs to some informal organization beginning to occur, though based more on need than on 'class-consciousness'.

In spite of the low unionisation rate of platform workers, some union confederations (SZEF, LIGA, MASZSZ - KASZ) explicitly insisted that *"there is a need to start to deal with this issue...."* The openness of the trade union leaders to representing platform workers depends on the sector/industry of their operation. For example, the LIGA-VISZ works in the hospitality sector is concerned with the problems of the AirBnb workers. MASZSZ – KASZ, representing the interests of the retail-sector workers, expressed interest in supporting all types of digital labour. LIGA showed a keen interest in becoming more active in issues relating to the digitisation of work.

Similarly, SZEF, which represents actors, musicians, members of the media sector, such as journalists, and camera operators who have been forced to move from the public to the private sector, often as self-employed KATA taxpayers, states that now is the time to start to deal with platform workers, such as small entrepreneurs or self-employees who are so-called 'KATA' taxpayers scheme and are experiencing financial problems.[6] The majority of platform workers are also KATA taxpayers, so they may be significantly affected by the current tax structure.

Under the impact of the COVID-19 pandemic, an increasing number of companies have begun implementing home office practice, which may continue to varying degrees after the pandemic. Finding the optimal balance between the life and work in this working arrangement is challenging, and it is likely that not only the home office will expand rapidly but self-employed platform work will as well. In this relation, it is necessary to note, that prior to the pandemic, among the 'CrowdWork21' project countries, the share of home-office workers in Portugal (39%) and Germany (38%) are above the European average (36%), while this type of working arrangement is less prevalent in Spain (30.5%) and in Hungary (28%) (Eurofound, 2020:5).

The problem of platform workers is related to other non-digital forms of employment too. For example, home care workers offering their services through on-line platform– mainly women – could work together in a cooperative form, guaranteeing also a kind of quality insurance.[7] For them, the union could provide advice, legal support and even financial help through a solidarity fund (SZEF).

---

[6] KATA - Itemised Tax for Small Businesses - is a flat tax so the accounting and deduction of coasts are not possible. If you are a full-time self-employed your KATA tax is 50.000 HUF or 75.000 HUF/month covering all corporate. The rules governing KATA will be modified from January 2021, the government wants to slap a 40% tax on KATA income over HUF 3 million, if the KATA taxpayer is invoicing a single company. This modification affects a lot of people, there are 377 000 KATA taxpayers in Hungary.

[7] Research suggests that service users pursue a cooperative model when the quality of cooperative services is perceived to be better than public, conventional private and non-profit alternatives (Cooperatives UK Limited. 2016; Vamsted, 2012). Contrary to such other care provider models, cooperatives do not simply administer services—they co-produce them (Conaty, 2014). Particularly in the

Union officials of MOSZ are more cautious. Despite their interest to take care of labour market participants, their position is that platform workers are out of reach for the classic trade unions. The head of secretariat at MOSZ (1) points out that in the case of platform workers, the trade union can find themselves in a schizophrenic situation: who should be protected, the platform or the platform workers? According to him, trade union must protect workers having clear employment relationship – easy to identify both employers and employees - and not those in a precarious status. There is a lack of motivation on the part of the union to organise low-status workers; the union will have only more work with them, without any income generated for them if they pay no dues or fees. *It is not the priority of interest of classical trade unions to represent the interests of the platform workers.*

Despite these difficulties, the unions surveyed raised their voice in the interest of precarious workers, but struggle or resist finding adequate strategies to address the needs of these workers. How, unions wonder, could organising and advocacy functions be combined for such workers? Information, legal aid services, and other basic support are among the essential roles of trade unions in relation to the care of platform workers. The classical trade union logic of recruiting union members is not applicable in a mechanistic way in the platform economy. Potentially, advocacy service could play a priority role in this growing workforce, although it will require creativity and renewal in the trade union communication and working structure. Innovative strategies, such as the use of a Facebook group, or other social media groups to diffuse and disseminate legal advice or organise charitable service for platform workers, will be needed to reach and support these workers.

The organisation of platform workers' interest – in contrast with the traditional trade union practice - could rather function as a social movement or association.

Among the trade union confederations/federations LIGA and LIGA - VISZ have the strongest motivation to engage with these new workers while continuing their efforts to represent the rights of unionised workers in the traditional economy. MOSZ shares the similar attitude, but SZEF and ÉSZT, representing public sector employees, and MASZSZ - KASZ, operating in the retail sector, have not prioritize the protection and interests of the already unionised workers.

---

multi-stakeholder model, users of care services become partners in care as voting members, rather than simply being recipients, working directly with care providers and staff to better target care plans." In: Providing Care through Cooperatives, Literature Review and case studies, ILO, 2017, p. 8.

According to MOSZ, the main motivation of the trade unions approaching platform workers would be to help them, first in providing legal assistance in the protection against digital platforms. The opinions differ in concern of 'protecting platform workers from the power of the digital platforms'. The two sectoral unions (LIGA – VISZ in the hospitality sector and MASZSZ – KASZ in the retail) and public sector's SZEF does not consider protection of platform workers as particularly relevant and important for them.

The evident lack of motivation or disinterest in platform workers may ultimately be a symptom of a deeper issue. According to the Hungarian Industrial Relations System, trade union membership is already low(9%), and yet with the exception of the MASZSZ – KASZ, it trade union confederations/federations admit they have no motivation to "recruit new members".

## 3.3 Unions' view on problems that platform workers experience and the need to implement a new regulatory (legal) framework

In the view of the trade unions, the main challenges that platform workers presently face are the following:

 1: Lack of social security, care and protection (LIGA, LIGA - VISZ, SZEF)
 2: Insecurity – lack of job security (LIGA, ÉSZT)

 3: High financial risk – unpredictable income (MASZSZ - KASZ)

 4: Labour law, legal protection (SZEF)

 5: Mental and physical exploitation, challenges of health and safety (ÉSZT)

According to LIGA – VISZ, which covers the hospitality industry, the main challenges for the workers in the platform economy are related with employment stability. In the context of the fast declining employment under the impact of the COVID-19 crisis, AirBnB platform workers realised that they are not entitled unemployment benefit or sick pay. Until the working culture and regulations of platform work are established, these workers must be considered at high risk in terms of employability and income security. According to MASZSZ-KASZ, platform workers who have self-employed status are prone to furthers risks (e.g. lack of sick pay, paid holiday, etc.), while SZEF observed that platform workers have no basic knowledge on the longer term dangers of this type of work (e.g. social security, employment stability, training, etc.). ÉSZT emphasized the further challenges of mental and physical risks and other health and safety related consequences of platform work.

In the opinions of the interviewed trade union leaders, the platform economy will likely continue to grow, and as a new model of future work and employment, trade unions have to be prepared for the related challenges. Elaborating on this view, the representative of SZEF felt that that the importance of this form of work will increase rapidly due to the COVID-19 crisis, a trend that is likely occurring in many labour markets around the world. However, due to the global and geographically dispersed nature of work, global platforms may easily avoid the attempts of national regulations. How will these workers - for example language teachers using the international teachers' platform - cope with the problems of accidents during work, welfare provision, issues of work-life balance, and so forth? Trade unions have to be well prepared if they commit to supporting these workers by, for example, developing appropriate training for platform workers to guarantee their long-term survival and adaptability on the on the online labour market (ÉSZT).

In the view of trade union leaders, the growth of the platform economy will be different in the various sectors. For example, the future employment of platform workers in the tourism industry (AirBnB) is highly uncertain due to the COVID-19 (LIGA - VISZ). According to MASZSZ – KASZ working in the retail sector, the biggest obstacle to the growth of the platform economy is that social and economic actors in the traditional economy have had substantial lobbying power in setting relevant legislation, preventing or obstructing the development of platform economy with laws and other form of regulation (MASZSZ – KASZ, see also Makó – Illéssy – Nostrabadi, 2020 on the case of Uber in Hungary).

*The consent based opinion of the trade union leaders/officials* on the regulation framework of platform work is that *the elaboration of legal regulation in the emerging platform economy is an essential requirement in order to avoid the generation an unfair competition between platforms and companies in the traditional economy*. There was general agreement that currently the Hungarian labour law legislation on the platform economy and platform workers is lagging behind its expansion (SZEF, MASZSZ KASZ). The rules in force are not suitable for handling the issues raised by the platform economy and/or workers and/or there are no rules at all (MASZSZ KASZ). The further problem is that the legislators themselves do not even know exactly what they should regulate. There is a need to study specific cases and debates (Rácz, 2020). Beside theoretical research, we need more empirically based ones.

Another huge challenge – according to the union leaders - is that the platform workers do not organise themselves nor recognize the need to do so. To protect their rights, they should inform themselves where the problems are (where 'do the shoes squeeze') and look for legal solutions (LIGA). LIGA - VISZ emphasizes the need for appropriate regulation in the field of labour regulation, health and safety and the field of hygienic standards. According to MOSZ (1), regulators dealing with platform economy over-concentrate on the issues of financial rules (e.g. taxation). In the light of international experiences, it is a rather narrow approach and can't cope with the complexity of challenges (e.g. working conditions, employment relations, skill development, etc.)

The crucial questions are the following: how the platform workers' interest could be protected and help to improve the quality of job as an important precondition for the sustainability of this form of work (MOSZ 2). Presumably, the durability of the COVID-19 crisis may speed of the diffusion of various types of digital work, including its platform-based version and may speed up efforts devoted to the much needed regulatory framework (MOSZ 1).

Transparency of the platform economy is another important pre-condition of the future regulatory framework. The profit-based platform economy should be transparent and not a kind of 'hidden' or 'grey' economy. Several unions ((LIGA, LIGA - VISZ) stressed that this is much-needed visibility is necessary to identify and address the inevitable violations of workers' rights. Working in the hospitality industry, LIGA – VISZ acknowledges that the appearance of such global platforms as AirBnB is favourable for service providers and service users (tourists), but it should not be allowed to operate in opacity in order to satisfy the platform workers aspirations too. In the employment praxis, the AirBnB working contracts are not transparent, and they are difficult to enforce for consumer and worker protection rules. Industry innovators and disrupters such as AirBnB and the governments that regulate them must accept that the platform economy has to co-exist or live side by side with the traditional economy and similarly has a strong global dimension too (SZEF).

### 3.4 Social dialogue in the platform economy: Important item on the 'wish-list'

Until recently, the common view of the Hungarian union representatives has been that *platform workers have not been involved into the process of social dialogue.* More specifically, ÉSZT has no any information on that issue. In the opinions of LIGA-VISZ and SZEF, a new operating structure for the social dialogue should be designed for the platform economy. The actual Sectoral Social Dialogue Committees (SSDC)[8] and the

---

[8] The interviewed trade union officials – when speaking on social dialogue – mentioned the role of the Sectoral Social Dialogue Committee (SSDC) in Hungary. The forums for sectoral social dialogue are the SSDC (in Hungarian: Szektoriális Társadalmi Dialógus Bizottság – ÁPB) supported by an EU programme (2003) and regulated by the law of LXXIV of 2009. They cover the sectors and employees under Labour Code. In these committees, the delegates of representative trade unions and employers' organisations have the right to negotiate agreements for the sector concerned. By now the activity of the ÁPBs has diminished since 2016 to almost to zero. The macro-level, national body for tripartite cooperation between workers' and employers' representatives and the government, the OÉT (National

representatives of social partners are unprepared, operating like a 'puppet show' (SZEF).

However, according to LIGA, SZEF and ÉSZT, it is possible and even necessary to establish social dialogue in the platform economy. LIGA stresses the need to change the legal regulation framework through, for example new labour laws, which includes the protection of self-employed, individual entrepreneurs and their right to negotiate.

SZEF stressed the need for innovation since there is a need to reform the established structure of social dialogue, which functions relatively well in the context of highly regulated (factory) work.. In the established industrial relations system, forums for social dialogue do exist, but the channels of information for participants are lacking (MOSZ 1). Developments related to the platform economy may give a new impetus to the renewal of industrial relations system in Hungary, and new solutions may appear in the next 5-10 years (MOSZ 2).

However, in the platform economy, it will be difficult for the social partners of labour relations to conduct social dialogue for the following reasons:

a) Platform activity isto a large degree – located into the 'grey' economy (LIGA - VISZ);

b) It is not clear who would be advocating for the employer (MASZSZ - KASZ); and

c) Platform workers are individual contractors (self-employed, freelancer, contract workers, etc.) and because they do not belong into an identifiable structure of the business organisation (company), it would be difficult to build up a community-based institution social relations (MASZSZ - KASZ).

Common view of the trade union leaders is that if there will be social dialogue for the platform workers the trade unions would have to play the role of interest representation

---

Interest Reconciliation Council – in Hungarian: Országos Érdekegyeztető Tanács, OÖT)), stopped to exist in 2012. Instead of it, a selected trade union confederations and employer organizations were invited by the government to establish the Permanent Consultation Forum (VKF) to consult on industrial policy. The actual official forum of *national level* social dialogue is the National Economic and Social Committee (NGTT), with the participation of not only of classical social partners of classical industrial relations system but also of business chambers, churches, etc.) and government as an observer. It is like an information and advisory forum without any right to decision. As ETHOS 2020 research states 'Most confederation and trade union leaders agree that social dialogue is more existent and alive on local level as there is more space left for maneuvering for the trade unions'. (ETHOS 2020, Sara Araújo & Maria Paula Meneses, The effectiveness of social dialogue as an instrument to promote labour and social justice, p.36. https://ethos-europe.eu/sites/default/files//docs/d6.4_website_report_complete.pdf).

using the internet infrastructure (through online platforms). LIGA even suggests that only the trade unions could represent the interests of the platform workers. According to SZEF, the new forum of social dialogue should differ from the present institution, which does not currently work properly.

The social dialogue designed for platform workers should be adapted to the specialities of their working arrangements (ÉSZT). LIGA-VISZ stresses that it must operate in the virtual space. Also according to MASZSZ-KASZ, as the present social dialogue has physical character, the new one for platform workers should be adapted to the employers' virtual presence and be carried out through online platforms.

The opinions of the trade union leaders/officials on the involvement of platform workers into the social dialogue vary on the potential for using the existing SSDC forum. According to the union confederations operating in the retail sector (MASZSZ – KASZ), the SSDC would be an 'ideal' forum to discuss the various aspects of working and employment conditions of platform workers. By contrast, according to MOSZ, the SSDCs of the traditional economy are not convenient for this purpose in their present form because the SSDC forum simply does not work.

In the opinion of LIGA, the legal framework for the freelancers and other platform workers must be created first in order to be included in the institution of social dialogue. According to SZEF, platform workers' bottom-up or grass-root organisation initiatives could be the solution. This will require strengthening the self-consciousness and increasing the agency of platform workers.

## 4. Challenging future roles for trade unions: advocacy vs. organising

According to the trade union representatives, there is no indication that a new trade union or other self- created or grass-root organisation of platform workers is being developed. (LIGA VISZ, LIGA, MOSZ, MASZSZ KASZ, SZEF, ÉSZT). LIGA and ÉSZT note that the main problem with recruiting platform workers or organising for them actions is that they are virtually invisible, they are hard to reach, trade unions do not know how to identify them, and they are geographically highly dispersed.

To get information on them and to organise them, national-level campaigns targeting these workers would be required. Due to the tensions between regulators and trade unions and the limited financial resources of trade unions, there is little chance such a campaign can be deployed (MASZSZ KASZ). If platform workers were to join a union, extra organisational and human capacity would be needed for their protection. However, according to the experiences of LIGA-VISZ, platform workers do not want to become trade union members. As platform workers are self-employed, it would be difficult to verify the basis of payment of their membership fee since for traditional unionized workers it deducted from their salary. According to MASZSZ-KASZ's experiences, those members who pay the membership fee independently (i.e. it is not deducted or transferred directly from the salary) eventually stop paying their membership fee. SZEF does not focus on this issue (recruiting platform workers) but if it did, it would likely struggle due to a lack of human resources and relevant knowledge on how to approach platform workers.

These substantial challenges aside, in theory unions could offer a wide range of support for platform workers:

> 1: In case of individual membership, unions could offer consultancy or legal advice on issues such as work contracts, legal representation, and labour law protection (LIGA, LIGA-VISZ, MASZSZ–KASZ, SZEF, ÉSZT), and social security registration (LIGA-VISZ);
>
> 2: Health and safety consultancy (ÉSZT);
>
> 3: Participation in trade union's training (MASZSZ–KASZ) or organisation of platform workers' training (SZEF, ÉSZT);
>
> 4: Knowledge transfer to become a conscious worker, organising meetings for them (SZEF);
>
> 5: Financial support for members in need (MASZ–KASZ), and possible creation of a solidarity fund (SZEF); and
>
> 6: Professional help from trade union experts to organise a union, should a certain group of platform workers would organise themselves to have a new union(LIGA).

LIGA - VISZ, representing the hospitality industry, have already consulted with employers' organisation on the topic of platform workers. They are also interested on related issues, such as complying with tax and social security regulation or fire safety, public health rules, etc., regulating activities of the traditional hotels. The other unions (LIGA, MASZSZ KASZ, MOSZ, SZEF, ÉSZT) have not yet established relations with the employers (platform owners) or public authority on the labour issues related with platform work. This issue is not currently at the forefront of the interest of national employers organisations. In contrast, the public authorities are increasingly interested in the issue of digitisation (e.g. Digital Welfare programme of the government, the special state secretariat at the Ministry of Innovation and Technology, which also deals with the platform economy.) The Ministry of Finances pays particular attention to the issues of financial regulations (e.g. taxation). There are conflicting opinions on the possible role of social dialogue - including their existing institutional form the SSDC - concerning platform workers. According to the sectoral union, MASZSZ - KASZ, SSDC could be a good place to discuss various work and employment related problems of platform workers. However, according to MOSZ, SSDC are not suitable for this purpose in their present form as they simply do not function effectively. Participants are missing and the sectoral logic does not fit for the dispersed and individualised nature of platform work (MOSZ 1).

*None of the interviewed unions has organised any campaigns towards digital workers thus far* (LIGA- VISZ, LIGA, MOSZ, SZEF, ÉSZT, MASZSZ - KASZ), and only MASZSZ - KASZ has planned to organise a campaign for warehouses and logistics workers taking part in online trade by the end of the summer (2020).Representatives of platform operating in the country have not attended any of the interviewed unions' meetings (LIGA VISZ, LIGA, MOSZ, MASZSZ - KASZ, SZEF, ÉSZT). *There is no information on the formation of any new trade union or other self-organisation of platform workers* (LIGA - VISZ, LIGA, MOSZ, MASZSZ - KASZ, SZEF, ÉSZT).

However, among the interviewed trade unions, some of them already received inquiries and contacted platform workers, but *these were always individual requests* (LIGA - VISZ, SZEF, MOSZ). In these cases trade union provided advice. ((LIGA VISZ, MOSZ, MASZSZ - KASZ). For example, platform workers approached LIGA - VISZ regarding their rights on wage and overtime payment. SZEF has already entered into contact with home caregivers and media workers. Sometimes the union leadership has not been supportive of these inquiries. The legal aid service of MOSZ used to receive individual questions from non-union workers, mostly on social security, taxation, invoicing or non-payment issues. However, they rarely receive questions on classic labour relation issues. MASZSZ - KASZ receives individual questions regularly, but not from platform workers. ÉSZT and SZEF have not yet received any inquiries from platform workers either.

MASZSZ - KASZ is open to any inquiry from individuals and is flexible in finding the right way toward further cooperation. SZEF emphasizes that the biggest problem for them is the lack of human resources to approach platform workers. They would need 2-3 persons and at least their one-year of effort to get any substantial result. At the ÉSZT, the issue to approach platform workers has not yet been raised.

According to all interviews, trade union leaders could and should control the organisation or representation of the platform workers. (LIGA, LIGA – VISZ, MOSZ, MASZSZ - KASZ, SZEF, ÉSZT).

**5. Contradictory trade union's views on the emerging grass-root organisations**

*The current situation is characterised by the complete lack of cooperation with grass-root initiatives* (LIGA - VISZ, ÉSZT, MOSZ, SZEF, LIGA, MASZSZ - KASZ). LIGA and MASZSZ - KASZ would cooperate if there were any new initiatives. MOSZ knows about some new associations but these groups are not yet in contact with the unions and try to move forward alone.

*The general opinion is that there are no organisations yet that would take the initiative to protect the platform workers' interests*, although this would be much needed (MOSZ 1, a MASZSZ - KASZ, SZEF, LIGA, LIGA - VISZ, ÉSZT). According to SZEF, the platform workers' bottom-up initiatives could be an appropriate solution for their interest representation. According to MOSZ (1), the issue of platform-workers' self-organisations is that they are still in their infancy stage in Hungary compared to Spain or Portugal. This can be attributed to the fact that in Spain the government supports such organisations, as does political and public communication. However, in Hungary, this support is lacking on the part of policy makers (MOSZ 3), allowing online platforms entrepreneurs (owners) to serve their own interests and not those of the platform workers.

The protection of platform workers in Hungary will not be solved overnight, and will need to be developed methodically and for the long term, building on grass-root initiatives, like 'Gólyások' (Storkers)[9]. which during the COVID19 pandemic tried to help bike couriers, and 'SMart'[10], which is taking care of freelancers, individual entrepreneurs, but not yet platform workers. This organisation protects the interests of those who join them, for example, if a customer does not pay for the service, they take it over and pay the worker for its service. , While there are differences in these two examples--behind the SMart there is business logic, while 'Gólya' is a non – profit organisation (MOSZ 1)—but such organizations are beginning to fill the vacuum that a union for platform workers could fill.

The role of the EU, according to LIGA, will be extremely important, in large measure because platform economy development has a strong global dimension (SZEF). European directives would help and speed up the creation of the much needed national regulation framework of the platform economy and platform workers (LIGA).

EFFAT-HOTREC, operating in the European hospitality sector and belonging to the European Trade Union Confederation (ETUC), is focusing on the working conditions, public health and consumer protection aspects of the platform economy and calls for the regulation at European level. Unfortunately, this process has now stopped (MASZSZ - VISZ), and because there are no rules at EU level yet, this allows strong anti-platform lobbies (like those against Uber) to prevail in the domestic markets (MASZSZ - KASZ). In this process, the role of ETUC is of great importance (SZEF).

---

[9] Gólya (Stork) was originally a cooperative pub and community house in Budapest VIII. district. In March 2020 they launched their courier service. See details: http://golyapresszo.hu/

[10] SMart was founded in 1998 in Belgium to support freelance artist, workers in creative industries to help them in admisitration and prevent financial unsecurity. They operate in several EU countries, including Hungary. See details: https://smarthu.org/

**6. Weak systematic use of social media by unions. Impacts of the COVID-19 crisis**

Opinions on the issue of social media are contrasting among the trade union confederations/federations. LIGA - VISZ, LIGA, SZEF, ÉSZT do not take part in social media group, this activity is not relevant for them. However, MASZSZ - KASZ is a member in several social media groups. The role of teleworking, distance learning, etc. is increasing (SZEF). However, the future of tourism-related platforms is uncertain; it depends not only on the vaccine but also on the possible end of the economic crisis triggered by the pandemic (LIGA - VISZ).

The interviewed union officials from LIGA - VISZ, LIGA and ÉSZT do not know of Facebook or WhatsApp groups where workers exchange their opinions or organise themselves. The representatives of SZEF and MOSZ are acquainted with Facebook groups but they do not have any function or activities regarding trade union coordination or any meaningful activity. According to MOSZ, there is a WhatsApp group with 5-600 members, and also, there are several groups on Facebook, such as for babysitters. However, their trade union coordination is zero, and trade unions are unable to develop any influence through them (MOSZ 1). SZEF knows about a Facebook group for home caregivers but without relevant activities. MASZ - KASZ created Facebook groups for their union members in order to improve communication among themselves and/or with the union[11].In addition, there are several companies in the retail sector where employees created their Facebook group.

According to LIGA - VISZ, LIGA, SZEF and ÉSZT, *presently there is no interest representative organisation (association/movement, owner or employer) which would control social media groups,* or at least they do not know about it. The social media groups connected to the unions are not controlled by the union (LIGA). The union control of the media groups is not relevant from the interest representation of the platform workers (LIGA). The Facebook groups generated within MASZSZ - KASZ

---

[11] https://www.facebook.com/KASZosok

https://www.facebook.com/groups/315217541877331 – Auchan KASZ; https://www.facebook.com/kasz.veszpremfejer; https://www.facebook.com/KASZ-N%C3%B3gr%C3%A1d-1006403059378110 – KASZ Nógrád are monitored by the staff of this trade union federation, and in addition this federation is also involved in content development. Besides, the Facebook groups created by the workers of a company – an international shopping mall - have an administrator supported by the trade union. In this case, it is quite typical that after a certain time, substantial information and content will appear which has nothing to do with the world of work. Typically, employers regularly monitor these pages to obtain information (MASZSZ - KASZ) to inform themselves of issues.

In the views of the unions leaders, the *impact of COVID-19* has a strong and complex impacts on digital working and particularly on platform workers.(7) For example, LIGA trade union confederation is preparing a comprehensive survey – covering 15 member unions – on the consequences of the pandemic on workers (but not specifically on platform workers). They wish to understand the conditions helping or inhibiting the survival of employment, how state aid was used, etc...

COVID-19 has affected some groups of platforms workers very substantially, especially in the hospitality industry. According to MOSZ (1) where traditional sectors are struggling due to pandemic, the platform companies are equally affected. Where the business has gone well despite the pandemic, its platform firms similarly prospered. LIGA-VISZ reported that many employees in their (hospitality) sector have lost jobs in both traditional hotels and catering and in its platform-based form: AirBnB. ÉSZT reported on major difficulties in the field of education, health and social care and public sectors. SZEF reported a major stop of culture and art related activities due to limitation in social distancing. MOSZ 2 draws the attention to an interesting new phenomenon: because of COVID-19, there has been an increase in the sensitivity and social solidarity towards the newly emerging platform workers, such as food couriers (Wolt, NetGo, etc.)

## 7. Concluding remarks and future challenges

The core experience learned from the interviews conducted among Hungarian trade union confederations/federations leaders/officials and from relevant research literature is a growing concern and uneasy situation on how to best cope with the challenges of interest representation in the fast growing digital economy. This is mainly due to the lack of basic financial, human and knowledge sources. Until now, Hungarian trade unions have not succeeded in systematically collecting empirical experiences on the working, employment conditions and on identifying and amplifying the collective voice (community) formation of platform workers or digital labour in general. In addition, they have been unable to fill this knowledge gap with the systematic and comparable empirical research produced by the academic community in Hungary and in other Central and Eastern European (CEE) countries. While our knowledge is constantly growing over time, the existing studies based on empirical evidences (both surveys and case studies) are not easily comparable, due to the lack of consent based terminology, research technics and the comprehensive sample of platform work and the countries.

To create a tailor-made and successful trade union strategy to organise platform workers is a daunting task to achieve because they constitute *a heterogeneous group of worker*s with different self-identities, bargaining positions, and differing needs. For example, to organise low-skilled, low-paid platform workers executing micro works and representing the essence of digital taylorism or digital precariat require quite another approach from traditional trade unions than those medium- or high-skilled, better paid freelancers (often calling themselves as self-entrepreneurs and representing medium class individuals) who execute complex, knowledge-intensive medium or highly specialised intellectual works. These more skilled workers have a much better bargaining position and do not care much about trade unions. Digital precariat, in contrast, represent the other extreme of the spectrum: their bargaining position is so weak that they often do not recognise how an interest representation organisation could help their everyday life. Even if they are (often extremely) dependent entrepreneurs, they do not have a distinct worker identity, and therefore they are hard to reach by traditional trade unions. This is where grass root organisation may come in and step up for their rights.

In an optimistic perspective, online work in general and platform work in particular may force unions to change the existing practice of their operations and discover new forms of recruiting new members in the context of the extremely low unionisation rate. For example, *online forums* are slowly emerging as important tools to articulate the collective voice of platforms workers who are often geographically dispersed and working alone. These 'community spaces' are supported by IT infrastructures of the platforms and '*external' forums,* such as Facebook groups or work blogs informing others of grievances and other issues, of platform workers. These forums may challenge and offer substitution of the existing (offline) institutions of collective voice. The international research experiences are illustrating the viability of this institutional renewal process, "… the ability for forums and online spaces to attract workers has led unions, worker centres, and other collective representation models to experiment with online forums and apps as part of a broader set of tools to assist in outreach and engagement." (Johnston – Land - Kazlauskas, 2019:31)

The experiences with workplace and employment related technological and organisational innovations call attention to the need to diversify or combine the trade union strategies in order to find the appropriate services for the fast growing new categories of workers. The traditional organisational tools in recruiting new members seem to us rather ineffective. There is need to identify the employee through recruiting methods that are consistent with the particular needs of the platform workers. For example, in addition to inventing new forms of recruiting techniques there is a need to focus more on the *strategy of advocacy* in contrast to the more traditional forms of organising strategies. During counselling, advising services could function as an organisational or collective learning process for both trade union staff and their new future 'clients' (various categories of the platform workers), while also creating mutual trust and engagement between trade unions and platform workers. Once *mutual trust and engagement* is created, it will be much easier to develop a shared vision and mutually reinforcing activities between workers and union organizers. The *cooperative based collective voice* "represents a distinctive approach for workers to achieve control in the workplace… They are being developed to provide services to gig workers …" (Johnston – Land – Kazlauskas, 2019:32).

The briefly outlined attempts to renew the existing institutions of collective voice or to create original new ones help identify challenges as well as possibilities for collective actions of labour (industrial) relations actors. The new working practices of platform work are not constructed and shaped in advance by technology. Their characteristics, working and employment relations are shaped by the by the labour (industrial) relations actors' value, social norms and preferences. Admittedly, the current situation does not seem to favour such a change. First, the *capital-labour nexus has dramatically changed* during the several past decades due to globalisation, and the global deregulation of the labour markets. As a consequence, a new regulation scheme has been developing from the 1970s, one which may lead to further dis-embeddedness of labour in the Polanyian sense (Montalban et al., 2019). Again, it is worth noting the highly segmented character of platform workers. This dis-embeddedness (i.e. detaching individuals from their social relations by further deregulation and by softening the rules governing these social relations) affects the groups of platform workers differently: it may favour the highly-skilled, highly-paid segment while further aggravating the subjection of the low-skilled, low-paid micro workers.

Second, the transformation of the Hungarian labour relations system shows that the role and weight of social dialogue has been decreasing at national and sectoral level in the past three decades. In the absence of formal employment relationship, these two levels of social dialogue are of crucial importance when it comes to interest representation of platform workers. In addition, *the declining institution of social dialogue* primarily affects negatively the trade unions that have over recent years, lost the majority of their membership and thus their bargaining power. *Scant financial, organisational and* *human resources* makes it extremely difficult to be present among platform workers and under these circumstances, it is understandable that organising the digital precariat – being a huge challenge even for more well-endowed trade unions – is not a top priority for the majority of Hungarian trade unions.

# References

Borbély, Sz. – Neumann, L. (2019) Neglected by the State: the Hungarian Experience, In: Müller, T. – Vandaelle, K. -. Waddington, J. (eds.) *Collective Bargaining in Europe: Towards and Endgame,* Vol. 4, Chpt. 14, pp. 297 – 314

Conaty, P. (2014) *Social co-operatives: A democratic co-production agenda for care services in the UK*. (Manchester, UK, Co-operatives UK

Cooperatives UK Limited. (2016) CASA: Social care with a difference. *Cooperatives UK Case Studies.* (Manchester, UK: Cooperatives UK Limited

Johnston, H. – Land-Kazlauskas, Ch. (2019) Organising on demande: Representation, Voice, and Collective Bargaining in the Gig Economy, *Conditions of Work and Employment Services,* No. 94, Geneva: ILO, p. 54

*Living, working and COVID-19. First findings – April 2020,* Dublin: European Foundation for Living and Working Conditions, p. 11

Maanen, J. V. (1979), "Reclaiming Qualitative Methods for Organizational Research: A Preface", *Administrative Science Quarterly,* Vol. 24 No. 4, pp. 520-526., In: Tomory, É. (2014) In: Tomory, É.M. (2014) *Bootstrap Financing: Case Studies of Ten Technology-Based Innovative Ventures*, Tales from the Best, Pécs: University of Pécs – Faculty of Business and Economics, PhD Dissertation, p. 134

Martinak, M. (2020) Interpretation of Amandement of Act no. 56/2012.Call. on Road Transport, Bratislava: *CEE Attorneys*, July, p. 2

Montalban, M. – Frigant, V. – Jullien, B. (2019) Platfrom Economy as a new form of capitalism: a Régulationist research programme, *Cambridge Journal of Economics,* Vol. 43 (4): 805-424doi: 10.1093/cje/bez017

Makó, Cs. – Illéssy, M. – Nostrabadi, S. (2020) The fiasco of Uber in Hungary, *EUWIN Newsletter,* Autumn, p. 3

Makó, Cs. (1997) From Monism to Divided Unionism, In: Schiensctock, G. – Thompson, P. – Traxledr, F. (eds.) *Industrial Relations between Command and Market,* New York: NOVA Scientific Publisher, pp. 24-43

Meszmann, T. (2018) Industrial Relations and Social Dialogue in the Age of Collaborative Economy (IRSDACE), National Report of Hungary, Bratislava*: Central European Labour Studies Institute (CELSI)*, December, p. 62

Nacsa, B. – Neumann, L. (2013): Hungary: The reduction of social democracy and employment. In: Lerais, Frédéric F. et. al. (eds): Social democracy under the strain of crisis. An essay of international comparison. Paris:, Institut de Recherches Économiques et Sociales (IRES). pp.), 94–108.

Neumann, L.(2018) A munka jövője – a szakszervezetek jövője? (Future of Work is the Future of Trade Unions?) *Magyar Tudomány (Hungarian Science),* 179. évf., No. 1. January, pp. 77 - 89.

Sedlakova, M. (2018) Industrial Relations and Social Dialogue in the Age of Collaborative Economy (IRSDACE), National Report: *Slovakia, Central European Labor Studies Institute, (CELSI)* p. 47, Available at: www.celsi.sk

Vamsted, J. (2012) Co-production and service quality: A new perspective for the Swedish welfare state. In: V. Pestoff, et al. (eds.), *New public governance, the third sector, and co-production*, New York: Routledge, p. 424, https://doi.org/10.4324/9780203152294

World Economic Forum (WEF) (2020) Charter of Principles for Good Platform Work. Geneva: WEF.